# Extraordinary Photostability and Davydov Splitting in BN-Sandwiched Single-Layer Tetracene Molecular Crystals

Seonghyun Koo[1], Ina Park[1], Kenji Watanabe[2], Takashi Taniguchi[2], Ji Hoon Shim[1,3] and Sunmin Ryu[1]*

[1]Department of Chemistry, Pohang University of Science and Technology (POSTECH), Pohang, Gyeongbuk 37673, Korea

[2]National Institute for Materials Science, 1-1 Namiki, Tsukuba 305-0044, Japan

[3]Department of Physics and Division of Advanced Materials Science, Pohang University of Science and Technology, Pohang (POSTECH), 37673, Republic of Korea

*E-mail: sunryu@postech.ac.kr

**ABSTRACT**

Two-dimensional molecular crystals have been beyond the reach of systematic investigation because of the lack or instability of their well-defined forms. Here, we demonstrate drastically enhanced photostability and Davydov splitting in single and few-layer tetracene (Tc) crystals sandwiched between inorganic 2D crystals of graphene or hexagonal BN. Molecular orientation and long-range order mapped with polarized wide-field photoluminescence imaging and optical second-harmonic generation revealed high crystallinity of the 2D Tc and its distinctive orientational registry with the 2D inorganic crystals, which were also verified with first-principles calculations. The reduced dielectric screening in 2D space was manifested by enlarged Davydov splitting and attenuated vibronic sidebands in the excitonic absorption and emission of monolayer Tc crystals. Photostable 2D molecular crystals and their size effects will lead to novel photophysical principles and photonic applications.

**Introduction**

Size and shape control of inorganic materials led to the discovery of quantum confinement effects in semiconducting nanosheets of $MoS_2$[1] in the 1960s and quantum dots of CuCl[2] and CdS[3] in the 1980s. Van Hove singularities[4] in carbon nanotubes[5] and massless linear dispersion[6] in graphene[7] are another manifestation of the dimensional effects and originate from the modified boundary conditions imposed on their electronic wave functions. Recent studies on 2-dimensional (2D) semiconductors revealed further details of the size effects such as the transition between direct and indirect bandgaps[8], strongly bound excitons[9] and trions[10], and inter- and intralayer redistribution of excitonic wavefunctions in 2D inorganic heterocrystals[11]. From the same perspective, 2D molecular crystals (2DMCs) can be a versatile platform to study dimensional effects on their crystalline structures and excitonic behaviors. The presence of proximate surfaces and modified dielectric screening in 2DMCs may induce structural changes like relaxation and reconstruction[12] observed in the surfaces of bulk solids. The structure-property relationship suggests a significant change in the electronic structures of 2DMCs that depend on intermolecular interactions. For example, the Davydov splitting (DS)[13] of photogenerated excitons in solid tetracene is greatly affected by the intermolecular interactions[14] and plays a decisive role in revealing their electronic structures. The exciton dynamics[15] in 2DMCs will be another key observable to be used in elucidating their electronic landscape and creating novel applications in energy and electronics.

To be a reliable material system for such photophysical and spectroscopic scrutiny, however, 2DMCs need to have sufficient long-range order and be stable under intense optical radiation. Thin films of organic molecules and polymers have long been studied for soft conductors[16], light-emitting diodes[17], thin-film transistors[18], and solar cells[19]. For example, the molecular films of polyacenes were either amorphous or polycrystalline, and typically varied from hundreds of down to several nm in thickness. Whereas conventional surface science methods allowed the growth of monolayer films on well-defined facets of noble metals[20-21] and hydrogen-passivated silicon[22], such samples were characterized mostly in-situ or incompatible with spatially-resolved optical spectroscopy and imaging. More recently, however, 2D molecular micro-films grown on inorganic layered crystals[23] showed crystallinity and compatibility with various ex-situ characterizations[24-26]. Given high quality and photostability, 2DMCs may serve as a general material platform to study the dimensional effects of molecular crystals.

In this work, we report quantum size effects in 2D tetracene (Tc) crystals by investigating their formation, structure and optical anisotropy of single and few layers grown on graphene and hexagonal BN (hBN). Encapsulation with graphene or hBN layers endowed 2D Tc, ambient-unstable otherwise, with greatly enhanced photostability. The DS of strongly polarized vibronic transitions in 2D Tc was substantially increased compared to 3D bulk crystals. Crystallographic domains and their preferential orientation with respect to the hBN substrates were also revealed by polarized wide-field emission imaging combined with optical second-harmonic generation (SHG) spectroscopy. Stable 2DMCs and their dimensional effects demonstrated in this work will play a pivotal role in exploring novel photophysical principles and photonic applications.

**Results**

**Growth, structure and encapsulation of 2D Tc crystals.** A triclinic bulk crystal of Tc belongs to the space group P$\bar{1}$ with the lattice constants of **a** = 7.98 Å, **b** = 6.14 Å, **c** = 13.57 Å, **α** = 101.3°, **β** = 113.2°, **γ** = 87.5°[27]. Its unit cell contains two basis Tc molecules in a herringbone configuration and spans a single Tc layer with the principal molecular axis almost perpendicular to the layer. Because both surfaces of each

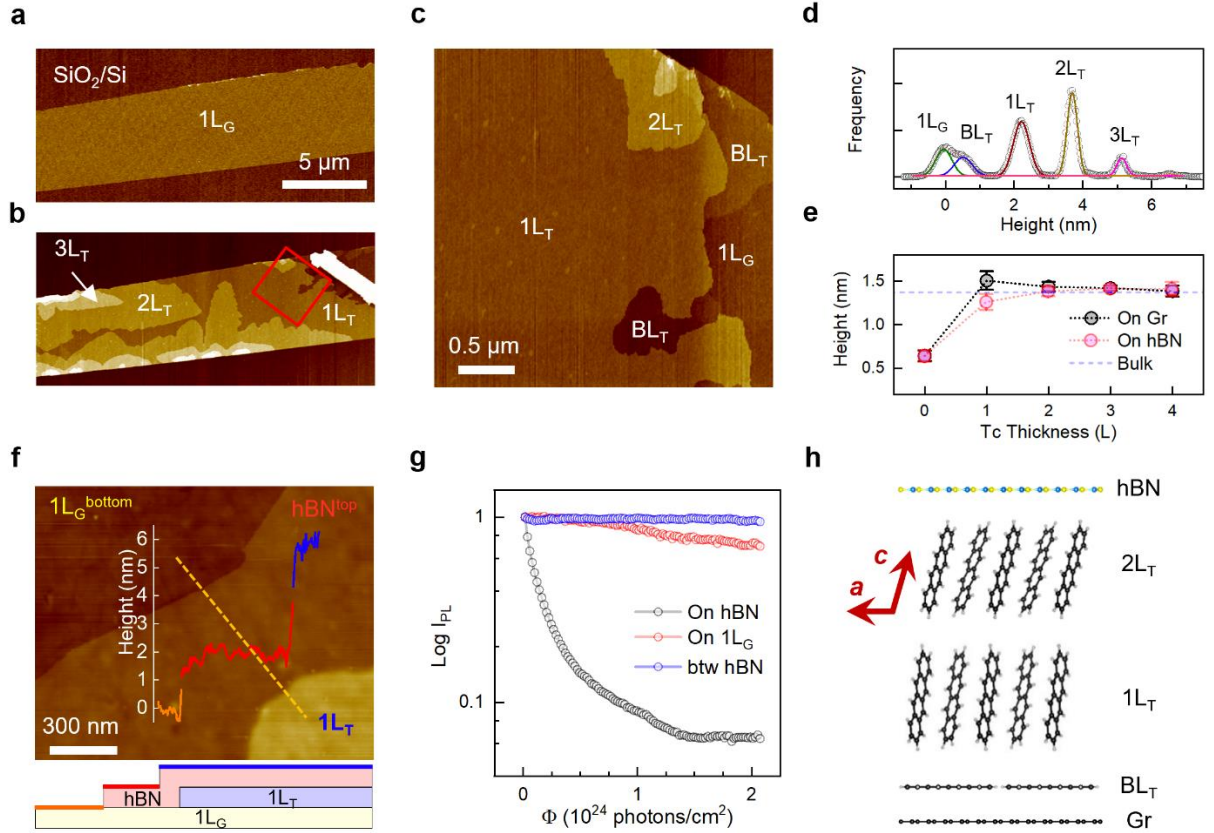

**Figure 1. Structure and photostability of 2D tetracene (Tc) crystals.** (a ~ c) Height images of 2D Tc crystals on $1L_G$ before (a) and after (b & c) deposition, respectively. The red-square area in (b) is given in (c). (d) Height profile obtained from (b) and (c). (e) Height of the topmost layer of $nL_T$ on graphene (Gr) and hBN as a function of Tc thickness (n), where n = 0 refers to $BL_T$. (f) Height image of confined $1L_T$ between the bottom ($1L_G$) and top (2 nm-thick hBN) walls. The height profile in yellow was obtained along the orange dashed line. (g) Photoinduced change in PL intensity of $1L_T$/hBN, $1L_T/1L_G$, and hBN/$1L_T$/hBN. The polarization of the excitation beam at 457 nm was aligned along the **b** axis of $1L_T$. Detailed analysis for photodesorption kinetics is given in Supplementary Note A. (h) Side view of hBN/$2L_T/1L_G$, where **a** and **c** denote the crystallographic axes of bulk Tc crystals.

Tc layer are terminated with H atoms, the interlayer adhesion is weaker than the intralayer bonding through π-π molecular interaction[28]. The layered nature of bulk Tc crystals presents the possibility of forming uniform 2DMCs of Tc. Indeed, activated self-assembly led to recrystallization of Tc into micron-wide layered films on 2D inorganic layered materials. Figures 1a and 1b showed the height topographies of single-layer graphene ($1L_G$) obtained with atomic force microscopy (AFM) before and after thermal evaporation of Tc at 40 ºC (see Methods). By varying the substrate temperature and the nominal thickness of Tc deposits, layered molecular films of Tc (denoted $nL_T$) could be formed. Depending on the context, $nL_T$ may refer to the $n^{th}$ Tc layer, not the whole film. Without thermal activation, less structured aggregates or thick rods were mostly generated (Fig. S1).

The layer-by-layer and dendritic growth starting from the edges of graphene suggest the crystalline nature[22, 29] of Tc films, which will be validated later. A closer look (Fig. 1c) revealed less conspicuous plateaus of buffer layers ($BL_T$) directly interacting with the substrates[25]. The height histogram and thickness profiles (Fig. 1d and 1e) indicated that the average thickness of a topmost Tc layer is $0.64 \pm 0.06$ nm for $BL_T$, reaches its maximum of $1.51 \pm 0.11$ nm ($1L_T$), and decreases to $1.38 \pm 0.06$ nm (thicker layers), which is equivalent to the interlayer spacing of bulk Tc crystals[27]. Whereas Tc also formed similar 2D structures including buffer layers on hBN substrates (Fig. S2), the thickness of a single layer gradually converged to the bulk value without a maximum (Fig. 1e). The presence of $BL_T$ and modulation in the single-layer thickness indicates that the balance between inter- and intralayer interactions varies for substrates and the overall thickness of Tc layers. We also deduced that the Tc molecules in $BL_T$ are mostly parallel to the substrates as observed on noble metals[20-21] from its reduced thickness. As depicted in the schematic side view in the **ac**-plane (Fig. 1h), the first and second π-stacked Tc layers form on $BL_T$ with varying tilt angle of the principal molecular axis. This geometric information based on Fig. 1e indicates that the interaction with the substrates plays a pivotal role in defining the structural details of 2D Tc crystals in competition with the π-π molecular interaction.

Despite the high melting point (357 °C) of bulk Tc, few-layer forms lasted only for several days in the ambient conditions because of desorption (Fig. S3). Moreover, even low-exposure laser irradiation for photoluminescence (PL) measurements was sufficient to disrupt Tc layers (Fig. S4). To enhance their stability, we encapsulated 2D Tc layers with graphene or few-layer hBN using the dry transfer[30] as shown in Fig. 1h. The height image in Fig. 1f indicated that the mechanical deposition of the hBN cover did not damage the $1L_T$ layer. The step height of Tc obtained after deposition of top covers typically increased by 0.5 ~ 2 nm,[31-32] which can be attributed to trapped molecules of ambient origin.[33] The sandwiched 2D Tc layers showed no noticeable changes even after 9-month storage in the ambient conditions (Fig. S5).

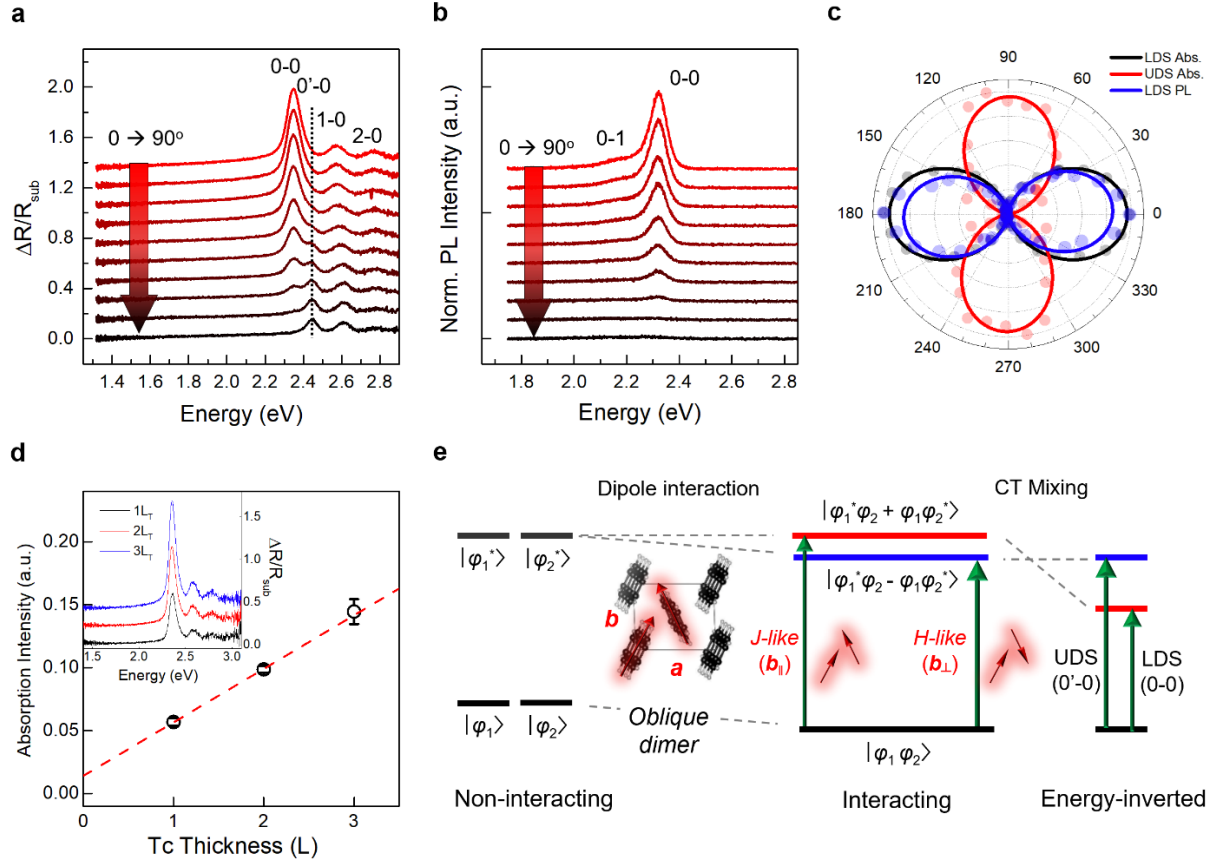

**Figure 2. Anisotropic light-matter interaction of 2D Tc crystals.** (a & b) Absorption by differential reflectance (a) and PL (b) spectra of $1L_T$/hBN obtained as a function of the polarization angle with respect to the **b** axis. Each of the vibronic peaks is designated with u-v, where u and v denote the vibrational level of the upper and lower electronic states, respectively. (c) Polar graph of absorption and PL intensity for 0-0 (LDS) and 0'-0 (UDS) transitions obtained from (a) & (b). The data were fitted with $\cos^2\theta$ (solid lines). (d) Area of the LDS absorption peak as a function of Tc thickness. The inset shows the b-polarized absorption spectra of $nL_T$ on $2L_G$. (e) Scheme for Davydov splitting (DS) in interacting oblique dimers energy-inverted by CT mixing, where $\varphi_n$ and $\varphi_n^*$ are the ground and excited electronic wave function of n-th basis molecule, respectively. Red and green arrows denote molecular transition dipole moments and allowed transitions to LDS and UDS, respectively.

Notably, they were also photo-resistant as shown in Fig. 1g that presents the PL intensity as a function of photon dose. The PL signals ($I_{PL}$) of $1L_T$ induced by the excitation at 457 nm consisted of fluorescence from the first excited state, as detailed in Fig. 2. Whereas $I_{PL}$ decreased drastically on hBN substrates without a cover, no hint of degradation was observed when protected with hBN covers (Fig. 1g). We also observed a substantial photostability on graphene substrates, unlike hBN (Fig. 1g). The post-irradiation AFM images (Fig. S4) suggest that the decrease in $I_{PL}$ was induced by photodesorption. A simple kinetic analysis led to an initial photodesorption cross section of $4.5 \times 10^{24}$ and $1.0 \times 10^{23}$ cm$^2$ for $1L_T$ on hBN and $1L_G$, respectively (Supplementary Note A). The pronounced suppression on graphene can be attributed to the efficient deactivation of excited Tc molecules, as shown by the 60-fold decrease of $I_{PL}$ on graphene (Fig. S6). Whereas the quenching process can be understood as Förster-type resonance energy transfer[34-35] as observed previously[36-37], the molecular details of the desorption process needs further investigation.

**Anisotropic light-matter interaction of 2D Tc crystals.** We show that 2D Tc is ordered and its optical response is highly anisotropic. In Fig. 2a, a series of polarized reflectance spectra of $1L_T$/hBN were obtained as varying the polarization angle of the incident light (Fig. S7). Differential reflectance of thin samples supported on transparent substrates is linearly proportional to the degree of absorption[38] (Methods). Notably, the reflectance (hereafter referred to as absorption) spectra exhibited a prominent vibronic progression with a spacing of 226 meV that is 15 meV larger than that of bulk Tc crystals[39]. The two peaks located respectively at 2.35 and 2.44 eV are induced by DS[13] of the 0-0 vibronic transition[14, 39] from $S_0$ to $S_1$. In the Kasha-type dimer model[40] described in Fig. 2e, the Coulombic interaction between the two transition dipole moments creates H and J-like Davydov states. Mixing of charge transfer (CT) character inverts the two[14], forming upper (UDS) and lower (LDS) Davydov states with their transition dipoles aligned perpendicular and parallel to the **b** axis, respectively[39]. The fact that both can be accessed by optical absorption indicates that the two basis Tc molecules form an oblique pair, unlike H or J-type dimers[40]. As depicted in Fig. 2e, we conclude that Tc molecules form a herringbone-arranged unit cell as observed in the bulk crystals.

As the polarization angle was varied, each of the 0-0 Davydov peaks showed distinctive modulation in its intensity (Fig. 2a). Their intensity profiles were well described by $\cos^2 \theta$ with clear nodes (Fig. 2c), indicating that both transitions are distinctively polarized and Tc molecules are well ordered. Notably, their transition dipole moments represented by the two-fold lobes formed an angle of $89.6 \pm 1.5°$. Their orthogonality validated the dimer model for the lowest exciton in 2D Tc crystals. Moreover, their **b** axes could be determined from the polarization direction of the LDS, as shown in Fig. 2c. Remarkably, the splitting for the 0-0 transition (95 meV) was 22% larger than that of bulk Tc, which is due to reduced screening effects, as will be discussed below. DS for the 0-1 and 0-2 transitions (Fig. 2a) was similar to that of bulk crystals, as summarized in Fig. S8.

The emission from 2D Tc is also polarized along the **b** axis and originates mostly from the LDS. The PL spectra in Fig. 2b showed a vibronic progression less pronounced than that observed in the absorption. The UDS emission was 200 times reduced considering its relative oscillator strength for absorption (Fig. S9) because of the efficient internal conversion from the UDS to LDS[41]. The Stokes shift for the 0-0 transition is 27 meV indicating minute relaxational effects[42]. The angular profile of the 0-0 emission intensity in Fig. 2c showed that the emitting transition dipole is parallel to that of absorption (full-angle spectra shown in Fig. S10). We also note that the emissive transition generating vibrationally excited Tc (denoted 0-v, v > 0) is 62% reduced when compared to aggregated or polycrystalline bulk samples (Fig. S11). In the excitonic coupling model based on the Frenkel-Holstein Hamiltonian[43], the fractional intensity of the vibrational sidebands is increased in the presence of structural disorder. This observation indicates superior structural order in the current 2D Tc crystals.

Notably, the LDS (UDS as well) absorptions from apparently monolithic $BL_T$ and $nL_T$ areas (Fig. S12) are polarized in the same direction and also well described with the $\cos^2 \theta$. This fact indicates long-range ordering possibly extended across the whole sample area and interlayer locking of orientation as in bulk crystals[27]. This interpretation is further supported by the fact that the peak intensity of LDS is linearly proportional to their number (n) of Tc layers (Fig. 2d). The absorption by $BL_T$ that corresponds to the intercept (0.014) of the linear fit is one-third of that for a single Tc layer given by the slope (0.043 per layer). The reduced absorption for $BL_T$ requires that its molecular number density per layer is significantly smaller than that for the upper layers, which will be theoretically validated below. The long-range order of $BL_T$ was also confirmed by the polarized emission shown in Fig. S13.

**SHG-assisted wide-field PL imaging of crystalline domains & polytypes.** The long-range order of 2D Tc was directly mapped with polarized PL imaging. Because the **b** axis is parallel to the polarization of the excitonic emission, the crystallographic orientation could be determined with a simple trigonometric manipulation of PL signals respectively obtained with the polarization of excitation beam in the horizontal and vertical configurations as marked in Fig. 3a and 3b (Supplementary Note B). The horizontal PL image in Fig. 3a exhibited a significant variation in its intensity ($I_H$), which mostly coincided with the thickness measured with AFM images and PL intensity (Fig. S14). However, the $2L_T$ area in the upper right corner (in blue circle) showed an unusually low intensity compared to other areas of the same thickness. Notably, the vertical PL image (Fig. 3b) revealed the opposite trend for its intensity ($I_V$). Figure 3c presents the spatial map of the azimuthal angle of the **b** axis ($\theta_b$), which was determined from the relation: $\theta_b = \tan^{-1}[(I_H/I_V)^{1/2}]$. Despite the wide-varying intensity maps, most of the 2D Tc areas are directed along one direction, whereas the upper right area points to another. The histogram for $\theta_b$ (Fig. 3d) provided a quantitative orientational distribution for the $hBN/1L_T/2L_G$ sample: 73% for 74 ± 6°, 20% for 64 ± 11°, and 6% for 15 ± 15°.

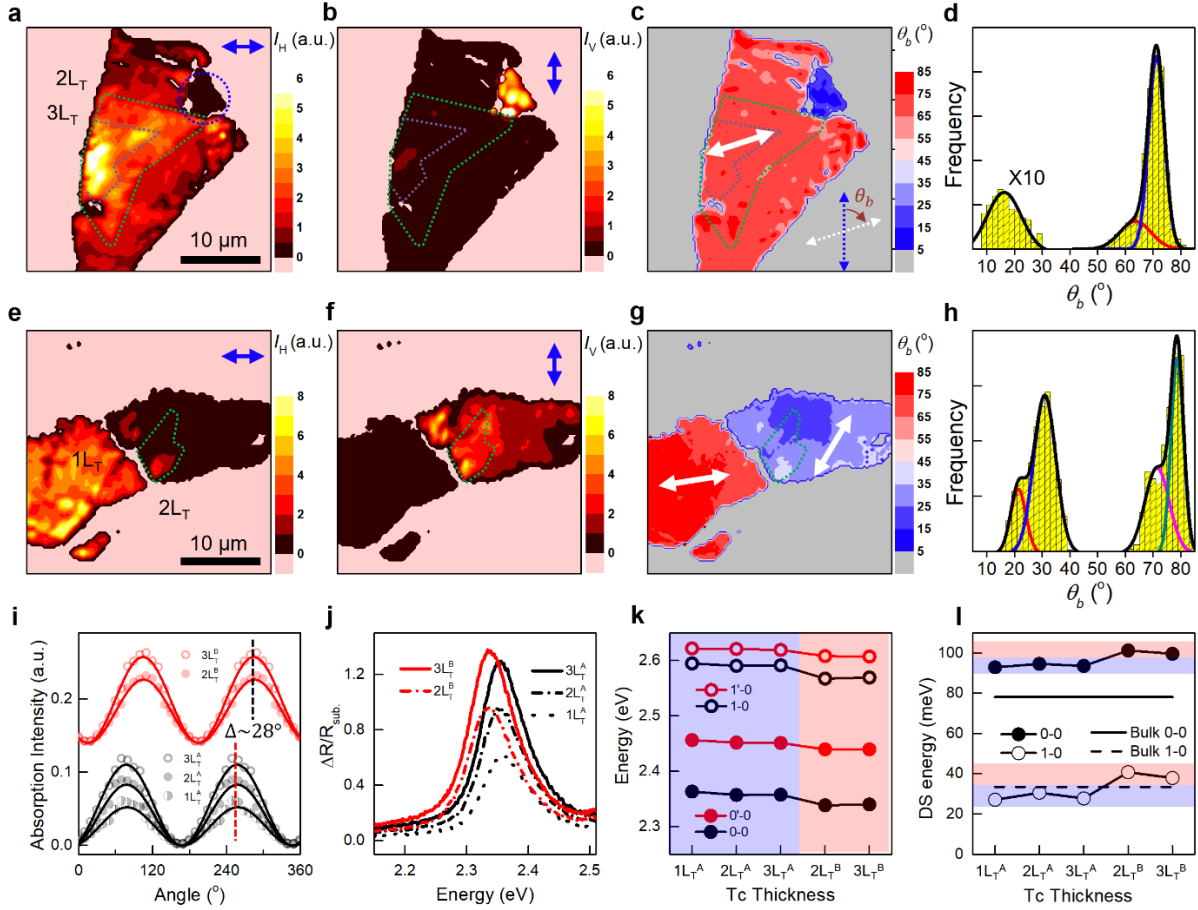

**Figure 3. Wide-field PL imaging of polytypes on graphene.** (a, b, e & f) Polarized PL intensity ($I_H$ & $I_V$) images of two hBN/nL$_T$/2L$_G$ samples, where blue arrows denote the polarization directions of the excitation beam at 457 nm: horizontal (a & e) and vertical (b & f). (c & g) Orientation ($\theta_b$) images for **b** axis (represented by white arrows) deduced from the $I_H$ and $I_V$ images, where $\theta_b$ is defined with respect to the vertical polarization (blue arrow). (d & h) $\theta_b$-histograms obtained from (c) and (g), respectively. Solid lines represent Gaussian fits. The data for 0 ~ 30° of (d) were multiplied by 10 in frequency for better visibility. (i) Absorption for LDS obtained as a function of polarization angle from two domains (nL$_T^A$ & nL$_T^B$) grown on single-crystal graphene (see Fig. S12 for morphology). (j) Representative spectra of (i) with maximum absorption for LDS. (k) Absorption energy of DS pairs for 0-0 and 1-0 transitions. The data were obtained from the two domains in (i). (l) DS energy of 0-0 and 1-0 transitions deduced from (k). Solid and dotted lines represent the bulk values of DS energy.

Whereas multiple Tc domains form on graphene as shown for another sample (Fig. 3e ~ 3h), no apparent orientational relation among them was found. Remarkably, the exciton energy varied significantly across Tc domains on graphene. Figure 3i presents the absorption polar graphs of two distinctive Tc domains

(denoted with superscript A and B). The $\cos^2\theta$-fit showed that the two domains are aligned with an angular offset of 28 ± 1°. The absorption spectra in Fig. 3j show that the LDS energy differs by 20 meV between the two domains, whereas their peak intensities are essentially equivalent. Note that the energy of the LDS exciton is constant within 2 meV for 2D Tc on hBN (Fig. S15). Then, the substantial energy differences for 0-0 and 1-0 (Fig. 3k) and DS (Fig. 3l) on graphene suggests that the detailed crystalline structure differs for the two Tc domains. It may originate from their unequal orientation with respect to the lattice of graphene, suggesting that nL$_T$/nL$_G$ exists in multiple polytypes or polymorphs of organic-inorganic heterocrystals.

In contrast, wide-field imaging revealed virtually unidirectional Tc growth on hBN, unlike on graphene, as shown in Fig. 4a ~ 4d (see Fig. S16 for additional examples). The formation of 2D Tc crystals spanning more than tens of microns implies their epitaxial registry with underlying substrates, as proposed in Fig. 4g. To determine the crystallographic orientation of hBN with respect to that of 2D Tc, we employed polarized SHG spectroscopy[31-32, 44]. During SHG, a non-centrosymmetric medium like hBN of odd-number layers converts two fundamental photons into one of doubled energy. The intensity of SHG signals (I$_{SHG}$) is dictated by the second-order susceptibility of the medium and its orientation with respect to the fundamental beam. For hBN belonging to the $D_{3h}^1$ space group, the parallel component of I$_{SHG}$ obeys an angular relation of $\cos^2 3\theta$ (see Fig. 4e and Methods for details). The polar intensity graphs in Fig. 4e were obtained from bare hBN areas of two 1L$_T$ samples as marked in Fig. S17. The intensity maxima of the 6-lobed patterns occur when the fundamental beam's polarization is parallel to the armchair directions (AC) of hBN. Then, the angular dependence of the absorption by the LDS superimposed in Fig. 4e indicates that the **b** axis of 2D Tc is aligned midway between two neighboring AC or almost parallel to zigzag directions (ZZ). The scheme in Fig. 4g presents a proposed epitaxial registry between hBN and 2D Tc. Assuming that the angle between **a** and **b** axes is the same as the bulk γ value of 87.5°, two equivalent stacking patterns can exist with their **a** axes parallel to AC as depicted in Fig. 4e. Notably, their **b** axes are offset from ZZ either by +2.5° or -2.5°, which agrees well with the findings in Fig. 4e. We conclude that Tc forms one type of 2D heterocrystals with hBN, unlike graphene.

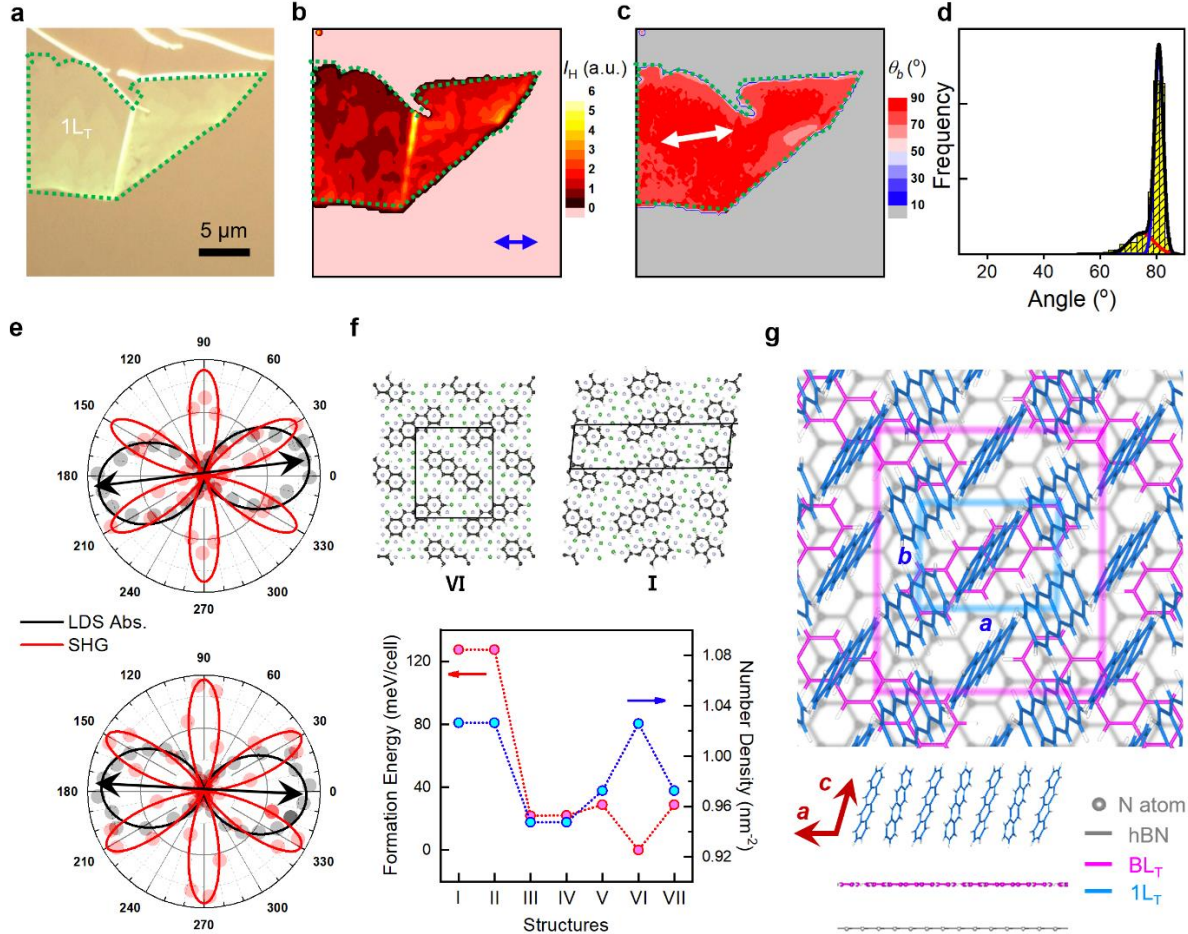

**Figure 4. Crystallographic registry between 2D Tc and hBN.** (a ~ c) Optical micrograph (a), $I_H$ (b), and $\theta_b$ (c) images of $nL_T$ on 4 nm-thick hBN. Green dotted lines delineate the area with Tc. (d) $\theta_b$-histogram obtained from (c). (e) Absorption (gray circles) and SHG (red circles) polar graphs obtained from two $1L_T$/hBN samples, where the zero angle corresponds to the ZZ direction of hBN and the black arrows represent the **b** axis of $1L_T$. (f) Structural models of the most (**VI**) and less (**I**) stable $BL_T$/hBN determined with first-principles calculations (top); Relative formation energy and number density of 7 $BL_T$/hBN structures (bottom). Green and gray balls in the models denote B and N atoms, respectively. See Supplementary Note C and Fig. S19 for details. (g) Proposed structure model of 2D Tc on hBN. Basis Tc molecules and unit cells of $BL_T$ and $1L_T$ are shown in magenta and blue, respectively. Whereas the orientations of both unit cells with respect to that of hBN were determined by polarized SHG and absorption spectroscopies, the structure of $BL_T$ and its registry with hBN were validated by first-principles calculations.

**Registry between organic and inorganic 2D crystals.** The above SHG analysis and the interlock between BL$_T$ and its upper layers (Fig. S12) led us to conclude that BL$_T$ is coupled to the underlying hBN with a distinctive atomic registry. To elucidate the molecular origin of the structural preference, we performed first-principles calculations on the adsorption of Tc on hBN in comparison with graphene (Supplementary Note C). Among various configurations, the face-on Tc molecule with its principal axis aligned along ZZ and their four hexagons centered at N atoms (denoted Tc$^{ZZ-N}$) is most stable as shown in Fig. S18, which agrees with the case of pentacene on hBN[25]. Remarkably, the calculation predicted that Tc$^{ZZ-N}$ is 81 meV larger in adsorption energy than Tc aligned along AC (denoted Tc$^{AC}$) and 127 meV than ZZ-aligned Tc with their four hexagons centered at B atoms (denoted Tc$^{ZZ-B}$). The energy difference between Tc$^{ZZ}$ and Tc$^{AC}$ on graphene was only 20 meV (Fig. S18a), which supports the presence of multiple orientational polytypes on graphene but not on hBN. Even a smaller energy difference was predicted on nanosized graphene platelets.[45]

To determine the most stable BL$_T$/hBN structure, we selected 9 two-basis arrangements consistent with the above experimental and theoretical results, as shown in Fig. S19. First of all, BL$_T$ consists of face-on Tc$^{ZZ-N}$ molecules (Fig. S18). The Davydov doublet in the polarized absorption (Fig. S12) requires a unit cell to contain two basis Tc$^{ZZ-N}$ molecules nearly in a herringbone pattern. The fact that the absorption of BL$_T$ is 32% of that for a single Tc layer (Fig. 2d) provided a rough estimate for the molecular number density of BL$_T$ (~1.3 nm$^{-2}$). We obtained 7 BL$_T$/hBN structures (Fig. S20) that satisfied the above requirements by placing the two basis molecules in close proximity (Supplementary Note C). The first-principles calculation showed that structure **VI** is most stable and its formation energy is ~120 meV/cell lower than that of **I** or **II** (Fig. 4f). Note that their two basis Tc molecules are aligned along ZZ so that their LDS transition dipole also points toward ZZ (Fig. 4e), as can be seen in the most and least stable BL$_T$ structures, **VI** and **I**, respectively (Fig. 4f). The fact that **VI** is one of the densest structures (Fig. 4f and Table S1) indicates that the stability of $BL_T$ is substantially dependent on intermolecular attraction in addition to vdW bonding between Tc and hBN.

**Discussion**

Our data showed that individual layers in 2D Tc are crystalline and its constituent molecules are flat-lying for $BL_T$ and almost edge-on for upper layers. The distinctive molecular interlock between $BL_T$ and hBN was found to originate from the ZZ-preferred adsorption of Tc. The fact that $BL_T$ and the upper layers are also aligned suggests a substantial interaction between the two moieties. The binding may be attributed to attractive quadrupolar interaction as found in T-shaped benzene dimers[46]. The most salient features in the absorption spectra of 2D Tc are the increased DS and decreased absorption by the vibrational sidebands compared to their bulk counterparts. The experimental DS (78 meV) of bulk Tc crystals[39] was an order of magnitude larger than expected in the Frenkel coupling scheme involving several excited singlet states but could be well reproduced by a refined model including the coupling between Frenkel and CT excitons[14, 47]. Thus the 22% enhancement in the DS of 2D Tc (Fig. 3l) suggests a stronger Frenkel-CT coupling or increased contribution of CT excitons. The excitons of 2D materials are strongly bound in general because of reduced screening[48], which should apply to 2D Tc and enhance Coulomb interactions, essentially stabilizing CT excitons. Theoretical simulations[14] predicted that the DS increases from 75 to 87 meV as the dielectric constant is reduced from 3 to 1. We also note that the decrease in the absorption by the vibrational sidebands is consistent with the reduced dielectric screening[14, 39].

**Conclusions**

We investigated the structure and excitonic behavior of Tc crystals in the 2D limit with polarized absorption and emission spectromicroscopy combined with SHG spectroscopy. Single and few-layer Tc crystals were formed with distinctive crystallographic orientation on hBN but randomly on graphene, the molecular origin of which was revealed by first-principles calculations. The buffer layers interlocks upper Tc layers with the underlying inorganic 2D crystals. The reduced dielectric screening in the 2D Tc crystals induced enlarged DS and attenuated vibronic sidebands compared to bulk Tc. Their instability in the ambient conditions was resolved by encapsulation with graphene or hBN, which led to their substantial resistance to intense laser beams. The geometric reinforcement and dimensional effects on the molecular excitons shown in this study can be achieved for other 2DMCs and instrumental in modifying their material properties for organic photonics and electronics.

## ASSOCIATED CONTENT

**Supporting Information.** Rodlike Tc aggregates, Morphology of 2D Tc crystals on hBN, Ambient instability of 2D Tc crystals, Photoinduced degradation of 2D Tc crystals, Enhanced stability of encapsulated 2D Tc crystals, PL quenching by graphene, Optical configuration of polarized confocal microscope, Analysis of Davydov pair of electronic transition, Insignificant emission from UDS, Full-angle spectra for Fig. 2a & 2b, Absorption and PL spectra of various Tc forms, Interlayer registry in $nL_T$ on $nL_G$ and hBN, PL spectrum of $BL_T$, Structure of encapsulated samples in Fig. 3, Consistency in PL energy of 2D Tc on hBN, Additional $nL_T$/hBN samples showing long-range order, Samples for polarized SHG and absorption measurements, Adsorption geometry and energy of single Tc molecule, Theoretical prediction of $BL_T$/hBN structure, Seven structural candidates for $BL_T$/hBN; Formation energy and Tc number density of 7 $BL_T$/hBN candidates; Determination of photodesorption cross-section of 2D Tc crystals, Imaging crystallographic orientation by ratiometric wide-field PL imaging, Theoretical prediction and experimental validation of structure of BLT/hBN. This material is available free of charge via the Internet at http://pubs.acs.org.

## AUTHOR INFORMATION

**Corresponding Author**

*E-mail: sunryu@postech.ac.kr

**Author Contributions**

S.R. conceived the project. S.K. and S.R. designed the experiments. S.K. performed the experiments and analyzed the data. I.P. and J.S. conducted the first-principles calculations. K.W. and T.T. synthesized hexagonal BN crystals. S.K. and S.R. wrote the manuscript with contribution from all authors.

**Notes**

The authors declare no conflict of interest.

**ACKNOWLEDGMENT**

S.R. acknowledges the financial support from Samsung Research Funding Center of Samsung Electronics under Project Number SSTF-BA1702-08. K.W. and T.T. acknowledge support from the Elemental Strategy Initiative conducted by the MEXT, Japan, Grant Number JPMXP0112101001 and JSPS KAKENHI Grant Number JP20H00354.

**REFERENCES**

1. Evans, B.; Young, P., Exciton Spectra in Thin Crystals. *Proc. R. Soc. A-Math. Phys. Eng. Sci.* **1967,** *298* (1452), 74-96.
2. Ekimov, A. I.; Efros, A. L.; Onushchenko, A. A., Quantum Size Effect in Semiconductor Microcrystals. *Solid State Commun.* **1985,** *56* (11), 921-924.
3. Rossetti, R.; Nakahara, S.; Brus, L. E., Quantum Size Effects in the Redox Potentials, Resonance Raman Spectra, and Electronic Spectra of CdS Crystallites in Aqueous Solution. *J. Chem. Phys.* **1983,** *79* (2), 1086-1088.
4. Van Hove, L., The Occurrence of Singularities in the Elastic Frequency Distribution of a Crystal. *Phys. Rev.* **1953,** *89* (6), 1189.
5. Wilder, J. W.; Venema, L. C.; Rinzler, A. G.; Smalley, R. E.; Dekker, C., Electronic Structure of Atomically Resolved Carbon Nanotubes. *Nature* **1998,** *391* (6662), 59-62.
6. Semenoff, G. W., Condensed-Matter Simulation of a Three-Dimensional Anomaly. *Phys. Rev. Lett.* **1984,** *53* (26), 2449.
7. Novoselov, K. S.; Geim, A. K.; Morozov, S. V.; Jiang, D.; Katsnelson, M. I.; Grigorieva, I. V.; Dubonos, S. V.; Firsov, A. A., Two-Dimensional Gas of Massless Dirac Fermions in Graphene. *Nature* **2005,** *438* (7065), 197-200.

8.  Splendiani, A.; Sun, L.; Zhang, Y.; Li, T.; Kim, J.; Chim, C. Y.; Galli, G.; Wang, F., Emerging Photoluminescence in Monolayer MoS$_2$. *Nano Lett.* **2010,** *10* (4), 1271-5.
9.  Qiu, D. Y.; da Jornada, F. H.; Louie, S. G., Optical Spectrum of MoS$_2$: Many-Body Effects and Diversity of Exciton States. *Phys. Rev. Lett.* **2013,** *111* (21), 216805.
10. Mak, K. F.; He, K.; Lee, C.; Lee, G. H.; Hone, J.; Heinz, T. F.; Shan, J., Tightly Bound Trions in Monolayer MoS$_2$. *Nat. Mater.* **2013,** *12* (3), 207-11.
11. Hong, X. P.; Kim, J.; Shi, S. F.; Zhang, Y.; Jin, C. H.; Sun, Y. H.; Tongay, S.; Wu, J. Q.; Zhang, Y. F.; Wang, F., Ultrafast Charge Transfer in Atomically Thin MoS$_2$/WS$_2$ Heterostructures. *Nat. Nanotechnol.* **2014,** *9* (9), 682-686.
12. Lee, C.; Yan, H.; Brus, L. E.; Heinz, T. F.; Hone, J.; Ryu, S., Anomalous Lattice Vibrations of Single- and Few-Layer MoS$_2$. *ACS Nano* **2010,** *4* (5), 2695-700.
13. Davydov, A. S., The Theory of Molecular Excitons. *Soviet Physics Uspekhi* **1964,** *7* (2), 145-178.
14. Yamagata, H.; Norton, J.; Hontz, E.; Olivier, Y.; Beljonne, D.; Bredas, J. L.; Silbey, R. J.; Spano, F. C., The Nature of Singlet Excitons in Oligoacene Molecular Crystals. *J. Chem. Phys.* **2011,** *134* (20), 204703.
15. Scholes, G. D.; Rumbles, G., Excitons in Nanoscale Systems. *Nat. Mater.* **2006,** *5* (9), 683-696.
16. Jérome, D.; Schulz, H.-J., Organic Conductors and Superconductors. *Adv. Phys.* **1982,** *31* (4), 299-490.
17. Gather, M. C.; Kohnen, A.; Meerholz, K., White Organic Light-Emitting Diodes. *Adv. Mater.* **2011,** *23* (2), 233-48.
18. Klauk, H., Organic Thin-Film Transistors. *Chem. Soc. Rev.* **2010,** *39* (7), 2643-66.
19. Cheng, Y. J.; Yang, S. H.; Hsu, C. S., Synthesis of Conjugated Polymers for Organic Solar Cell Applications. *Chem. Rev.* **2009,** *109* (11), 5868-923.
20. Huang, H.; Song, F.; Lu, B.; Zhang, H. J.; Dou, W. D.; Li, H. Y.; He, P. M.; Bao, S. N.; Chen, Q.; Zhou, W. Z., Coverage Dependence of the Structure of Tetracene on Ag(110). *J. Phys.-Condens. Matter* **2008,** *20* (31), 315010.
21. Smerdon, J.; Bode, M.; Guisinger, N.; Guest, J., Monolayer and Bilayer Pentacene on Cu(111). *Phys. Rev. B* **2011,** *84* (16), 165436.
22. Tersigni, A.; Shi, J.; Jiang, D. T.; Qin, X. R., Structure of Tetracene Films on Hydrogen-Passivated Si(001) Studied Via STM, AFM, and NEXAFS. *Phys. Rev. B* **2006,** *74* (20), 205326.
23. Wang, Q. H.; Hersam, M. C., Room-Temperature Molecular-Resolution Characterization of Self-Assembled Organic Monolayers on Epitaxial Graphene. *Nat. Chem.* **2009,** *1* (3), 206-11.
24. Zhao, H.; Zhao, Y.; Song, Y.; Zhou, M.; Lv, W.; Tao, L.; Feng, Y.; Song, B.; Ma, Y.; Zhang, J.; Xiao, J.; Wang, Y.; Lien, D. H.; Amani, M.; Kim, H.; Chen, X.; Wu, Z.; Ni, Z.; Wang, P.; Shi, Y.; Ma, H.; Zhang, X.; Xu, J. B.; Troisi, A.; Javey, A.; Wang, X., Strong Optical Response and Light Emission from a Monolayer Molecular Crystal. *Nat. Commun.* **2019,** *10* (1), 5589.
25. Zhang, Y. H.; Qiao, J. S.; Gao, S.; Hu, F. R.; He, D. W.; Wu, B.; Yang, Z. Y.; Xu, B. C.; Li, Y.; Shi, Y.; Ji, W.; Wang, P.; Wang, X. Y.; Xiao, M.; Xu, H. X.; Xu, J. B.; Wang, X. R., Probing Carrier Transport and Structure-Property Relationship of Highly Ordered Organic Semiconductors at the Two-Dimensional Limit. *Phys. Rev. Lett.* **2016,** *116* (1), 016602.
26. Amsterdam, S. H.; Stanev, T. K.; Zhou, Q.; Lou, A. J.; Bergeron, H.; Darancet, P.; Hersam, M. C.; Stern, N. P.; Marks, T. J., Electronic Coupling in Metallophthalocyanine-Transition Metal Dichalcogenide Mixed-Dimensional Heterojunctions. *ACS Nano* **2019,** *13* (4), 4183-4190.
27. Robertson, J. M.; Sinclair, V.; Trotter, J., The Crystal and Molecular Structure of Tetracene. *Acta Crystallogr.* **1961,** *14* (7), 697-704.

28. Northrup, J. E.; Tiago, M. L.; Louie, S. G., Surface Energetics and Growth of Pentacene. *Phys. Rev. B* **2002**, *66* (12), 121404.
29. Glicksman, M. E.; Lupulescu, A. O., Dendritic Crystal Growth in Pure Materials. *J. Cryst. Growth* **2004**, *264* (4), 541-549.
30. Castellanos-Gomez, A.; Buscema, M.; Molenaar, R.; Singh, V.; Janssen, L.; van der Zant, H. S. J.; Steele, G. A., Deterministic Transfer of Two-Dimensional Materials by All-Dry Viscoelastic Stamping. *2D Mater.* **2014**, *1* (1), 011002.
31. Ryu, Y.; Kim, W.; Koo, S.; Kang, H.; Watanabe, K.; Taniguchi, T.; Ryu, S., Interface-Confined Doubly Anisotropic Oxidation of Two-Dimensional $MoS_2$. *Nano Lett.* **2017**, *17* (12), 7267-7273.
32. Kim, W.; Ahn, J. Y.; Oh, J.; Shim, J. H.; Ryu, S., Second-Harmonic Young's Interference in Atom-Thin Heterocrystals. *Nano Lett.* **2020**, *20* (12), 8825-8831.
33. Haigh, S. J.; Gholinia, A.; Jalil, R.; Romani, S.; Britnell, L.; Elias, D. C.; Novoselov, K. S.; Ponomarenko, L. A.; Geim, A. K.; Gorbachev, R., Cross-Sectional Imaging of Individual Layers and Buried Interfaces of Graphene-Based Heterostructures and Superlattices. *Nat. Mater.* **2012**, *11* (9), 764-767.
34. Swathi, R. S.; Sebastian, K. L., Resonance Energy Transfer from a Dye Molecule to Graphene. *J. Chem. Phys.* **2008**, *129* (5), 054703.
35. Raja, A.; Montoya Castillo, A.; Zultak, J.; Zhang, X. X.; Ye, Z.; Roquelet, C.; Chenet, D. A.; van der Zande, A. M.; Huang, P.; Jockusch, S.; Hone, J.; Reichman, D. R.; Brus, L. E.; Heinz, T. F., Energy Transfer from Quantum Dots to Graphene and $MoS_2$: The Role of Absorption and Screening in Two-Dimensional Materials. *Nano Lett.* **2016**, *16* (4), 2328-33.
36. Xie, L. M.; Ling, X.; Fang, Y.; Zhang, J.; Liu, Z. F., Graphene as a Substrate to Suppress Fluorescence in Resonance Raman Spectroscopy. *J. Am. Chem. Soc.* **2009**, *131* (29), 9890.
37. Chen, Z.; Berciaud, S.; Nuckolls, C.; Heinz, T. F.; Brus, L. E., Energy Transfer from Individual Semiconductor Nanocrystals to Graphene. *ACS Nano* **2010**, *4* (5), 2964-8.
38. Mak, K. F.; Sfeir, M. Y.; Wu, Y.; Lui, C. H.; Misewich, J. A.; Heinz, T. F., Measurement of the Optical Conductivity of Graphene. *Phys. Rev. Lett.* **2008**, *101* (19), 196405.
39. Tavazzi, S.; Raimondo, L.; Silvestri, L.; Spearman, P.; Camposeo, A.; Polo, M.; Pisignano, D., Dielectric Tensor of Tetracene Single Crystals: The Effect of Anisotropy on Polarized Absorption and Emission Spectra. *J. Chem. Phys.* **2008**, *128* (15), 154709.
40. Hestand, N. J.; Spano, F. C., Expanded Theory of H- and J-Molecular Aggregates: The Effects of Vibronic Coupling and Intermolecular Charge Transfer. *Chem. Rev.* **2018**, *118* (15), 7069-7163.
41. Tayebjee, M. J.; Clady, R. G.; Schmidt, T. W., The Exciton Dynamics in Tetracene Thin Films. *Phys. Chem. Chem. Phys.* **2013**, *15* (35), 14797-805.
42. Lakowicz, J. R., *Principles of Fluorescence Spectroscopy*. 3rd ed.; Springer US: Singapore, 2006; p 954.
43. Spano, F. C., Modeling Disorder in Polymer Aggregates: The Optical Spectroscopy of Regioregular Poly(3-Hexylthiophene) Thin Films. *J. Chem. Phys.* **2005**, *122* (23), 234701.
44. Li, Y.; Rao, Y.; Mak, K. F.; You, Y.; Wang, S.; Dean, C. R.; Heinz, T. F., Probing Symmetry Properties of Few-Layer $MoS_2$ and h-BN by Optical Second-Harmonic Generation. *Nano Lett.* **2013**, *13* (7), 3329-33.
45. Zarudnev, E.; Stepanian, S.; Adamowicz, L.; Karachevtsev, V., Noncovalent Interaction of Graphene with Heterocyclic Compounds: Benzene, Imidazole, Tetracene, and Imidazophenazines. *ChemPhysChem* **2016**, *17* (8), 1204-12.
46. Hobza, P.; Selzle, H.; Schlag, E., Floppy Structure of the Benzene Dimer: ab initio Calculation on the Structure and Dipole Moment. *J. Chem. Phys.* **1990**, *93* (8), 5893-5897.

47. Beljonne, D.; Yamagata, H.; Bredas, J. L.; Spano, F. C.; Olivier, Y., Charge-Transfer Excitations Steer the Davydov Splitting and Mediate Singlet Exciton Fission in Pentacene. *Phys. Rev. Lett.* **2013,** *110* (22), 226402.
48. Chernikov, A.; Berkelbach, T. C.; Hill, H. M.; Rigosi, A.; Li, Y.; Aslan, O. B.; Reichman, D. R.; Hybertsen, M. S.; Heinz, T. F., Exciton Binding Energy and Nonhydrogenic Rydberg Series in Monolayer $WS_2$. *Phys. Rev. Lett.* **2014,** *113* (7), 076802.

**TOC Graphic**

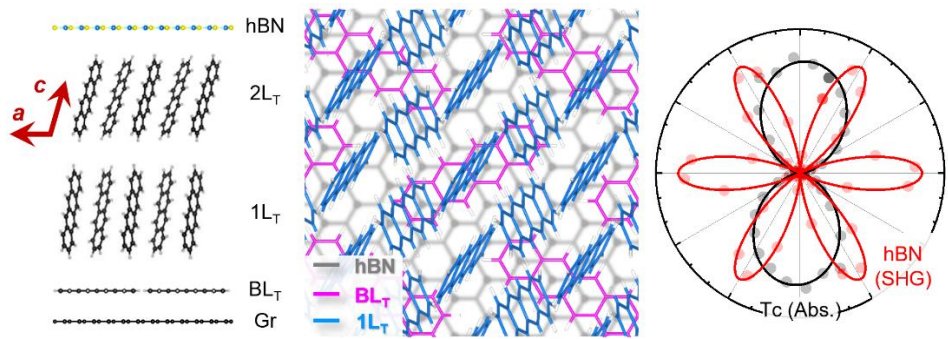